\documentclass[aps,prb,twocolumn]{revtex4}
\usepackage{graphicx,amsmath}
\usepackage[english]{babel}

\begin{document}
\title{Anisotropy of microwave conductivity of YBa$_2$Cu$_3$O$_{7-x}$
in superconducting and normal states: Crossover $3D-2D$}
\author{M.R.~Trunin, Yu.A.~Nefyodov}
\affiliation{Institute of Solid State Physics of RAS, 142432
Chernogolovka, Moscow region, Russia}

\begin{abstract}
Imaginary part of the microwave conductivity $\sigma''(T<T_c)$ and
the resistivity $\rho(T)=1/\sigma(T>T_c)$ along ($\sigma_{ab}''$
and $\rho_{ab}$) and perpendicular to ($\sigma_c''$ and $\rho_c$)
cuprate $ab$-planes of YBa$_2$Cu$_3$O$_{7-x}$ crystal were
measured in the temperature range $5\leq T \leq 200$~K, having
varied the oxygen doping level $x$ in the crystal from 0.07 to
0.47. The superconducting state was found to have the dependences
$\sigma_{ab}''(T)/\sigma_{ab}''(0)$ and
$\sigma_c''(T)/\sigma_c''(0)$ coincide in the optimally doped
($x=0.07$) crystal. With the increase of $x$ the dependences
$\sigma_c''(T)/\sigma_c''(0)$ have smaller slope at $T<T_c/3$ with
$\sigma_{ab}''(T)/\sigma_{ab}''(0)$ changing slightly. The
$ab$-plane transport in the normal state of YBa$_2$Cu$_3$O$_{7-x}$
crystal always remains metallic. However, a crossover from Drude
conductivity (at $x=0.07$) along the $c$-axis to the hopping one
(at $x>0.07$) takes place. This is proved by both estimating
minimum metallic $c$-conductivity and maximum tunneling
$c$-conductivity and quantitatively comparing the measured
dependences $\rho_c(T)$ with the ones calculated in the polaron
model of quasiparticles' $c$-transport.

\end{abstract}
 \maketitle

Recently a lot of interest has been attracted by investigations of
transport properties evolution in high-temperature superconductors
(HTSC) with different level of doping by oxygen and other
substitutional impurities; in other words, the dependence upon the
number $p$ of holes per Cu atom in the CuO$_2$ plane. The $p$
value and the critical temperature $T_c$ of a superconducting
transition in HTSC satisfy the following empirical relationship
\cite{Tal1}: $T_c=T_{c,max}[1-82.6(p-0.16)^2]$.

Up to now, the narrow band of HTSC phase diagram, corresponding to
the optimal doping ($p\approx 0.16$) and maximum values of the
critical temperature $T_c=T_{c,max}$, is the most studied area. In
the normal state of optimally doped HTSC the resistivity
$\rho_{ab}(T)$ of cuprate $ab$-planes increases proportionally to
temperature, viz. $\Delta\rho_{ab}(T)\propto T$. The resistivity
$\rho_c(T)$ in perpendicular direction substantially exceeds the
value of $\rho_{ab}(T)$ and also has metallic behavior (both
$\rho_{ab}(T)$ and $\rho_c(T)$ increase with $T$). Bi-2212, which
is the most anisotropic HTSC (having $\rho_c/\rho_{ab}\approx
10^5$ at $p\approx 0.16$), represents an exception: its
resistivity $\rho_c(T)$ increases as $T$ approaches $T_c$
($d\rho_c(T)/dT<0$). This Bi-2212 feature agrees with the estimate
of the minimal metallic conductivity in the $c$-direction for
anisotropic three-dimensional ($3D$) Fermi liquid model
\cite{Xie}:
\begin{equation}
\sigma_{c,min}^{3D}=\sqrt{\rho_{ab}/\rho_c}\,ne^2d^2/h,
\end{equation}
where $n\approx 10^{21}$~cm$^{-3}$ is a $3D$ carrier
concentration, $d$ is lattice constant along the $c$-axis, $h$ is
Planck's constant. In Bi-2212 the conductivity
$\sigma_c=1/\rho_c\ll \sigma_{c,min}^{3D}$ at $T=T_c$, but the
other optimally doped HTSC have $\sigma_c(T_c)>
\sigma_{c,min}^{3D}(T_c)$. The conductivity $\sigma_{c,min}$
in~(1) is less than Ioffe-Regel limit $\sigma_{IR}=e^2k_F/h$ for
the two-dimensional ($2D$) case, i.e., $\sigma_{c,min}\approx
\sqrt{\rho_{ab}/\rho_c}\,\sigma_{IR}\,d/a\ll \sigma_{IR}$
($a\approx 2\pi/k_F$ is the lattice constant in the CuO$_2$
plane), whereas $\sigma_{ab,min}\approx \sigma_{IR}$ \cite{Xie}.

A measure of HTSC anisotropy in the superconducting state is the
ratio
$\sigma_{ab}''(0)/\sigma_c''(0)=\lambda_c^2(0)/\lambda_{ab}^2(0)$,
where $\sigma_{ab}''$ and $\sigma_c''$ are imaginary parts of
conductivity, $\lambda_{ab}$ and $\lambda_c$ are penetration
depths of high-frequency field for currents running in the
$ab$-planes and perpendicular to them, correspondingly. It is
common knowledge that $\Delta\lambda_{ab}(T)\propto T$ at
$T<T_c/3$ in optimally doped high-quality HTSC and this
experimental fact provides strong evidence for $d_{x^2-y^2}$
symmetry of the order parameter in these materials. However, there
is no consensus in literature about $\Delta\lambda_c(T)$ behavior
at low temperatures. Even YBa$_2$Cu$_3$O$_{6.95}$ ($T_c\approx
93$~K), the most thoroughly studied single crystals, have shown
both linear, $\Delta\lambda_c(T)\propto T$ \cite{Mao,Srik,Nef1},
and quadratic dependences \cite{Hos} in the range $T<T_c/3$.

Pseudogap states, appearing at $p<0.16$, occupy a wide band of
HTSC phase diagram, which is far less investigated. Measurements
of $ac$-susceptibility of oriented HTSC powders at $T<T_c$ show
\cite{Xiang} that their dependences $\sigma_c''(T)/\sigma_c''(0)$
have smaller slope at $T\to 0$ than
$\sigma_{ab}''(T)/\sigma_{ab}''(0)$ do. The normal state of
underdoped HTSC is characterized by non-metallic behavior of the
resistivity $\rho_c(T)$ at $T$ approaching $T_c$, by deviation of
the resistivity from its linear dependence
$\Delta\rho_{ab}(T)\propto T$ and by $\rho_c/\rho_{ab}$ ratio
rising dramatically with the decrease of $p$. A lot of theoretical
models have been proposed to explain these properties, but there
is none fully describing the evolution of the dependences
$\sigma_{ab}''(T)$, $\sigma_c''(T)$ and $\rho_{ab}(T)$,
$\rho_c(T)$ in the wide range of concentration $p$ and temperature
$T$. Moreover, the $c$-transport mechanism is not defined; in
particular, it is not clear if it can be metallic (Drude), or at
any $p$ conductivity along the $c$-axis is determined by
quasi-particles tunneling between cuprate layers, accompanied by
their scattering both within the layers and between them.
\begin{table*}[t]
\vspace{-2mm}\caption{Annealing temperatures and parameters
characterizing doping and the anisotropy of the superconducting
and normal state of YBa$_2$Cu$_3$O$_{7-x}$.}\vspace{-3mm}
{
\vspace{0.5cm}
\begin{tabular}{|c|c|c|c|c|c|c|c|c|}

\hline annealing&critical&\multicolumn{2}{c|}{doping}&
\multicolumn{2}{c|}{$\lambda$ values}&$\Delta\lambda_c(T)\propto
T^\alpha$
&$\lambda_{c}/\lambda_{ab}$&$\sqrt{\rho_c/\rho_{ab}}$\\
temperature&temperature&\multicolumn{2}{c|}{parameters}&
\multicolumn{2}{c|}{at $T=0$}&{$\alpha$}&{at}&{at}\\
$T$, $^\circ$C&$T_c$, K&$p$&$x$&$\lambda_{ab}$, nm&$\lambda_{c}$, $\mu$m&&$T=0$&$T=200$K \\

\cline{1-9}
500&92&0.15&0.07&152&1.55&1.0&10&11\\
\cline{1-9}
520&80&0.12&0.26&170&3.0&1.1&18&18\\
\cline{1-9}
550&70&0.105&0.33&178&5.2&1.2&29&16\\
\cline{1-9}
600&57&0.092&0.40&190&6.9&1.3&36&16\\
\cline{1-9}
720&41&0.078&0.47&198&16.3&1.8&83&35\\
\hline
\end{tabular}}\end{table*}

The present paper analyzes the results of anisotropy measurements
and evolution of temperature dependences of conductivity
components in the YBa$_2$Cu$_3$O$_{7-x}$ crystal under varying
oxygen doping in the range $0.07\le x\le 0.47$. The crystal of a
rectangular shape, with dimensions $1.6\times 0.4\times
0.1$~mm$^3$, has been grown in a BaZrO$_3$ crucible. The
measurements were made at the frequency of $\omega/2\pi=9.4$~GHz
and in the temperature range $5\le T\le 200$~K. To change an
oxygen content in the sample, we successively annealed the sample
in the air at different $T\ge 500^{\circ}$~C specified in Table~1.
Anisotropy was measured for each of the five crystal states.
According to susceptibility measurements at the frequency of
100~kHz, superconducting transition width amounted to 0.1~K in the
optimally doped state ($x=0.07$), however, the width increased
with the increase of $x$, having reached 4~K at $x=0.47$. The
temperatures of the superconducting transition were $T_c=92, 80,
70, 57, 41$~K. The whole cycle of the microwave measurements
included the following: (i) we measured the temperature
dependences of the quality factor and of the frequency shift of
the superconducting niobium resonator with the sample inside in
the two crystal orientations with respect to the microwave
magnetic field, transversal ($T$) and longitudinal ($L$); (ii)
measurements in the $T$ orientation gave the surface resistance
$R_{ab}(T)$, reactance $X_{ab}(T)$ and conductivity
$\sigma_{ab}(T)$ of the crystal cuprate planes in its normal and
superconducting states; (iii) measurements in the $L$ orientation
gave $\sigma_c(T)$, $X_c(T)$, $R_c(T)$. See \cite{Nef1} for the
details of the measuring technique in the optimally doped
YBa$_2$Cu$_3$O$_{6.95}$ crystal. We have also reported the
temperature dependences of the surface impedance components of
YBa$_2$Cu$_3$O$_{7-x}$ at various $x$ in the short communication
\cite{Nef2}.

\begin{figure}[b]
\centerline{\includegraphics[width=0.49\textwidth,clip]{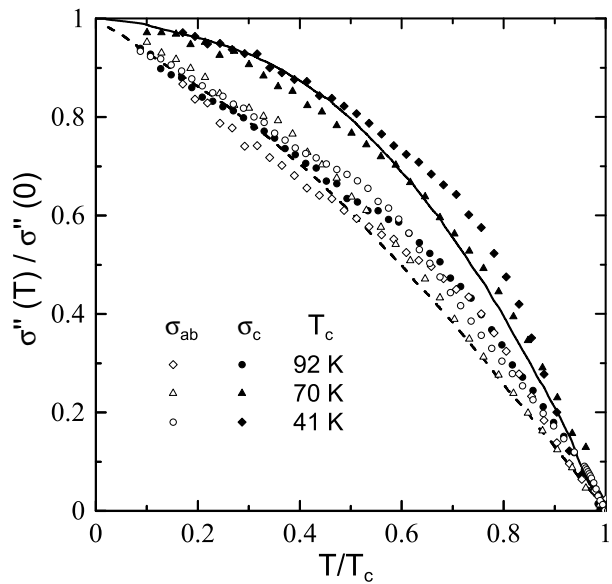}}\vspace{-3mm}
\caption{The dependences $\sigma_{ab}''(T)/\sigma_{ab}''(0)$ (open
symbols) and $\sigma_c''(T)/\sigma_c''(0)$ (full symbols) measured
for the three states of the YBa$_2$Cu$_3$O$_{7-x}$ crystal with
$T_c=92$~K, $T_c=70$~K and $T_c=41$~K. Solid and dashed lines
stand for the dependences $\sigma_c''(T)/\sigma_c''(0)$ and
$\sigma_{ab}''(T)/\sigma_{ab}''(0)$ calculated in~\cite{Lev1} for
YBa$_2$Cu$_3$O$_{7-x}$ with oxygen deficit.} \label{f1}
\end{figure}

Fig.~1 shows the dependences $\sigma_{ab}''(T)/\sigma_{ab}''(0)$
(open symbols) and $\sigma_c''(T)/\sigma_c''(0)$ (solid symbols)
at $T\le T_c$ for YBa$_2$Cu$_3$O$_{7-x}$ states with $T_c=92$~K,
$T_c=70$~K and $T_c=41$~K. Table~1 contains values of the
penetration depths $\lambda_{ab}(0)$ and $\lambda_c(0)$ at $T=0$.
The temperature behavior of the dependences
$\sigma_{ab}''(T)/\sigma_{ab}''(0)$ does not show dramatic change
at varying $p$. The optimally doped state of
YBa$_2$Cu$_3$O$_{6.93}$ features good matching of the temperature
dependences $\sigma_{ab}''(T)/\sigma_{ab}''(0)$ and
$\sigma_c''(T)/\sigma_c''(0)$. Only does the theory of the linear
response of an anisotropic superconductor \cite{Xiang} explain
this observation. With decreasing $p$ the temperature dependences
$\sigma_c''(T)/\sigma_c''(0)$ at $T<T_c/3$ become substantially
weaker than $\sigma_{ab}''(T)/\sigma_{ab}''(0)$.

The model \cite{Lev1} is appropriate for comparison with the
experimental data provided in the present paper. The model
describes the following contributions to the $c$-transport of
quasi-particles in the superconducting and normal HTSC states: (a)
direct hopping between cuprate planes and (b) hopping accompanied
by inelastic scattering on phonons, as well as elastic scattering
on impurities lying between planes. The conductivity in the very
cuprate planes is considered to be a Drude one
\begin{equation}
\sigma_{ab}=\frac{e^2\nu_{2D}D_{ab}}{d}=\frac{n_{2D}e^2\tau}{md},
\end{equation}
where $\nu_{2D}=m/\pi\hbar^2$ is the $2D$ density of states,
$D_{ab}=v_F^2\tau/2$, $v_F$, $\tau$ and $n_{2D}=k_F^2/2\pi$ are
the diffusion coefficient, Fermi velocity, relaxation time and the
$2D$ density of quasi-particles in the $ab$-plane,
correspondingly. The total Hamiltonian of the electronic system in
the model \cite{Lev1} represents a sum of Hamiltonians of $m$
individual CuO$_2$ layers, $\sum_m H_m$, and an inter-layer
coupling Hamiltonian $H_{\bot}$, which is considered small in
comparison with $\sum_m H_m$. This resulted in the perturbation
theory calculations to the second order of $H_{\bot}$ showing that
the quasi-particles transport between neighboring weakly-coupled
layers is analogous to tunneling in the $SIS$ junction at $T<T_c$
and in the $NIN$ junction at $T>T_c$. In addition, in the process
(a) the electron $ab$-plane momentum is conserved (specular
tunneling), whereas in the process (b) is not (diffusive
tunneling) \cite{Legg}.

\begin{figure}[t]
\centerline{\includegraphics[width=0.35\textwidth,clip]{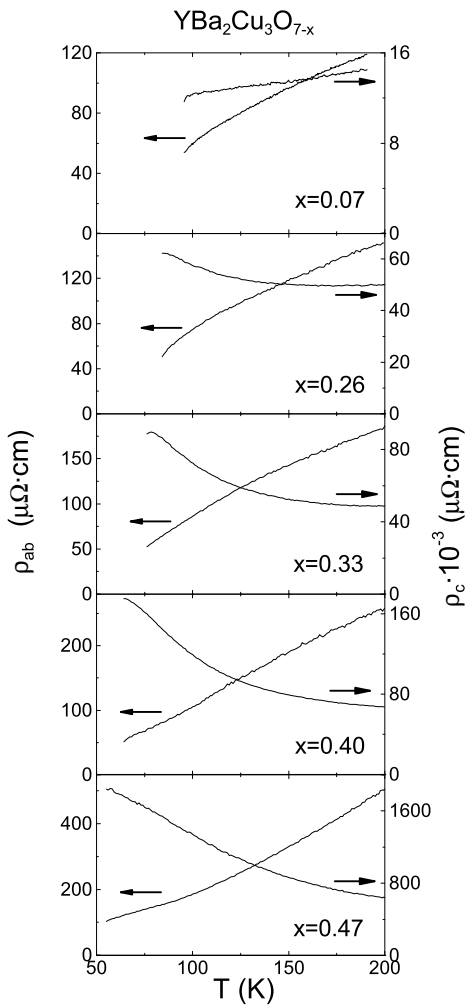}}\vspace{-3mm}
\caption{The evolution of the measured $\rho_{ab}(T)$ and
$\rho_c(T)$ dependences in YBa$_2$Cu$_3$O$_{7-x}$ with different
oxygen content.} \label{f2}
\end{figure}

The authors of \cite{Lev1} used the BCS model with $d$-symmetry of
the order parameter in CuO$_2$ layers to calculate anisotropy of
the superconducting HTSC state. Fig.~1 represents numerical
results, calculated taking into account both of the processes (a)
and (b), for $\sigma_c''(T)/\sigma_c''(0)$ (solid line) and for
$\sigma_{ab}''(T)/\sigma_{ab}''(0)$ (dashed line). Having compared
this with the experimental data at $T<T_c/2$ for
YBa$_2$Cu$_3$O$_{7-x}$, with oxygen deficiency $x>0.07$, one can
see dramatic decrease in the slopes of the dependences
$\sigma_c''(T)/\sigma_c''(0)$ with the increase of $x$, as well as
insignificant change of the dependences
$\sigma_{ab}''(T)/\sigma_{ab}''(0)$. The larger slopes of the
experimental dependences in comparison with the theoretical ones
at $T>T_c/2$ may be due to the effects of a strong electron-phonon
interaction \cite{Tru1}, which are not taken into account in the
model \cite{Lev1}. The dashed line in the Fig.~1 also coincides
with the dependence $\sigma_c''(T)/\sigma_c''(0)$, calculated in
\cite{Lev1} in the absence of diffusive tunneling (b), when the
specular tunneling process (a) along the $c$-axis becomes
identical to the $c$-transport in the $3D$ superconductor. This
exceptional case corresponds to the optimally doped
YBa$_2$Cu$_3$O$_{6.93}$.

\begin{figure}[b]
\centerline{\includegraphics[width=0.4\textwidth,clip]{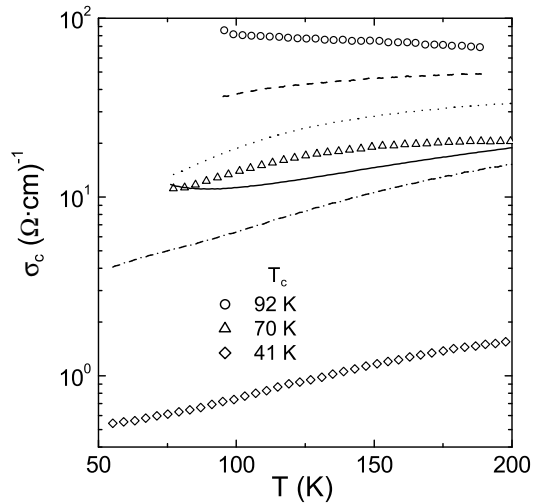}}\vspace{-3mm}
\caption{The symbols show for the experimental $\sigma_c(T)$
dependences in three states of YBa$_2$Cu$_3$O$_{7-x}$ at
$T_c=92$~K, $T_c=70$~K and $T_c=41$~K. The dashed, dotted and
dash-dotted lines stand for the values of
$\sigma_{c,min}^{3D}(T)$, corresponding to these states and
obtained from the formula (1) using the measured $\rho_{ab}(T)$
and $\rho_c(T)$ in Fig.~2. The solid line represents calculations
of $\sigma_c(T)$ using the formulas from~\cite{Lev1} for
YBa$_2$Cu$_3$O$_{6.67}$.} \label{f3}
\end{figure}

The real and imaginary parts of the surface impedance of the
YBa$_2$Cu$_3$O$_{7-x}$ crystal, measured for each value of $x$ in
the Table~1, were found to coincide at $T>T_c$ \cite{Nef2}:
$R_{ab}(T)=X_{ab}(T)$, $R_c(T)=X_c(T)$. Therefore, the
resistivities $\rho_{ab}(T)$ and $\rho_c(T)$ were found from
$R_{ab}(T)$ and $R_c(T)$, applying the usual formulas of the
normal skin effect: $\rho_{ab}(T)=2R_{ab}^2(T)/\omega \mu_0$,
$\rho_c(T)=2R_c^2(T)/\omega \mu_0$. Fig.~2 shows the evolution of
the dependences $\rho_{ab}(T)$ and $\rho_c(T)$ in the range
$T_c<T\le 200$~K with the change of $x$. The last column in
Table~1 contains $(\rho_c/\rho_{ab})^{1/2}$ values at $T=200$~K.
Only the optimally doped YBa$_2$Cu$_3$O$_{6.93}$ shows that both
dependences $\rho_{ab}(T)$ and $\rho_c(T)$ have a metallic
behavior, and the ratio $\rho_c/\rho_{ab}$ approaches the
anisotropy of effective masses of charge carriers
$m_c/m_{ab}=\lambda_c^2(0)/\lambda_{ab}^2(0)$ in the pure $3D$
London superconductor, which type YBa$_2$Cu$_3$O$_{6.93}$ belongs
to. The other states of YBa$_2$Cu$_3$O$_{7-x}$ with lower
concentration of holes have the resistivity $\rho_c(T)$ increase
with the decrease of temperature, which shows its non-metallic
behavior. Fig.~3 represents the experimental dependences
$\sigma_c(T)$ compared to the values of $\sigma_{c,min}^{3D}$,
calculated using the formula (1) for the three states of the
YBa$_2$Cu$_3$O$_{7-x}$ crystal with $T_c=92$~K (dashed line),
$T_c=70$~K (dotted line) and $T_c=41$~K (dash-dotted line). It is
only the $c$-axis conductivity of YBa$_2$Cu$_3$O$_{6.93}$ that
exceeds the minimum metallic value of $\sigma_{c,min}^{3D}$ in the
whole temperature range.

Thus, it is natural to assume that as well as in the
superconducting state of YBa$_2$Cu$_3$O$_{7-x}$, a slight decrease
in the carriers concentration (in comparison with the optimal
level) in the normal state results in the crossover from $3D$
metallic conductivity to $2D$ Drude conductivity in CuO$_2$ layers
and tunneling conductivity between these layers (the crossover
$3D$-$2D$). To analyze this assumption, the model \cite{Lev1}
again proves to be appropriate. If $t_{\bot}$ is the interplane
hopping matrix element, the $c$-conductivity of quasi-particles in
the process (a) will be as follows \cite{Lev1,Legg,Kum,Lar}:
\begin{equation}
\sigma_c^{dir}=2e^2\tau\nu_{2D}\left(\frac{t_{\bot}}{\hbar}\right)^2=
4\sigma_{ab}\left(\frac{t_{\bot}d}{\hbar v_F}\right)^2,
\end{equation}
where $2\tau(t_{\bot}/\hbar)^2$ is the intensity of the direct
tunneling between neighboring CuO$_2$ layers, $\sigma_{ab}$ is the
conductivity (2) in these layers. For this case the typical
hopping time $\hbar/t_{\bot}$ significantly exceeds the relaxation
time $\tau$ in the plane \cite{Legg}: $\hbar/t_{\bot}\gg\tau$. The
opposite limit $\hbar/t_{\bot}\ll\tau$ corresponds to Drude
conductivity in all three directions, as in an anisotropic $3D$
metal. When $\hbar/t_{\bot}\approx\tau$, a crossover takes place.
The maximal value of tunneling $c$-conductivity
$\sigma_{c,max}^{dir}=2\sigma_{IR}\sqrt{\rho_{ab}/\rho_c}$ from
(3) is then reached, which is approximately equal to the minimal
metallic conductivity $\sigma_{c,min}^{3D}$ from (1). In the case
of diffusive tunneling of quasi-particles (processes (b) in the
model \cite{Lev1}), conductivity along the $c$-axis equals
\cite{Legg,Graf}
\begin{equation}
\sigma_c^{diff}=\frac{e^2\nu_{2D}D_c}{d}=\frac{e^2\nu_{2D}d}{\tau_c},
\end{equation}
where $D_c=d^2\tau_c$ is the diffusion coefficient and
$1/\tau_c$ is the probability of scattering between cuprate
planes. As in the previous case, at $\tau_c\approx \tau$ we find
$\sigma_{c,max}^{diff}=\sigma_{IR}\sqrt{\rho_{ab}/\rho_c}\approx
\sigma_{c,min}^{3D}$, and from (2) and (4) we obtain another
representation of the $3D$-$2D$ transition criterion:
\begin{equation}
\sigma_{c,max}\sigma_{ab}\approx\frac{n_{2D}}{\pi}\left(\frac{e^2}{\hbar}\right)^2.
\end{equation}

At $n_{2D}=n/d\approx 10^{14}$~cm$^{-2}$ the formula (5)
demonstrates that the crossover $3D$-$2D$ appears when the value
of $\rho_c\rho_{ab}\approx 10^{-6}$~($\Omega\cdot$cm)$^2$ is
reached. The data in the Fig.~2 gives $\rho_c\rho_{ab}\lesssim
10^{-6}$~($\Omega\cdot$cm)$^2$ at $x=0.07$ only, which confirms
that the $3D$ anisotropic Fermi-liquid model is applicable to
explaining the properties of the optimally doped
YBa$_2$Cu$_3$O$_{6.93}$.

Two temperature dependences, with distinction of kind, of the
$c$-conductivities at $T\ge T_c$ are followed from the formulas
(3) and (4): for the case of the direct tunneling,
$\sigma_c^{dir}(T)\propto\sigma_{ab}(T)$ rises with the increase
of $\tau(T)$ at $T$ approaching $T_c$, while $\sigma_c^{diff}(T)$
falls with the increase of $\tau_c(T)$. According to the model
\cite{Lev1}, the total conductivity $\sigma_c$ along the $c$-axis
equals the sum of the conductivities, which are in turn determined
by each of the above processes, (a) and (b). In the vicinity of
$T_c$, it is scattering of quasi-particles on impurities, lying
between cuprate planes, that determines $\sigma_c^{diff}$, which
is not temperature dependent, as long as another contribution to
$\sigma_c^{diff}$, arising due to interaction with phonons,
vanishes with temperature decrease. Vice versa, at $T\gg T_c$ the
phonon contribution becomes prevailing. Thus, we have the
following approximate temperature dependence for the conductivity
$\sigma_c(T)$: $A/T+C+BT$ ($A, B, C$ are temperature independent),
which does not allow for our experimental data (the solid line in
Fig.~3 represents an example of calculating $\sigma_c(T)$ using
formulas from \cite{Lev1} for the sample YBa$_2$Cu$_3$O$_{6.67}$.
\begin{figure}[b]
\centerline{\includegraphics[width=0.44\textwidth,clip]{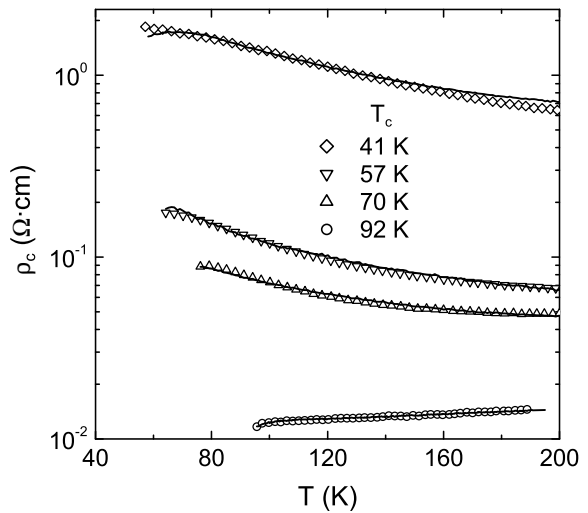}}\vspace{-3mm}
\caption{Comparison of the experimental dependences $\rho_c(T)$ in
YBa$_2$Cu$_3$O$_{7-x}$ (symbols) and those calculated from (6)
(solid lines).} \label{f4}
\end{figure}

However, very recently another $c$-transport model was proposed
\cite{Ho}, well describing all the dependences $\rho_c(T)$ shown
in Fig.~2. Unlike \cite{Lev1}, where the effects of interaction
with phonons appear in the second order of the perturbation
theory, in the model \cite{Ho} the Hamiltonian allows for these
effects accurately through a canonical transformation \cite{Lan};
it is only after this interplanar tunneling of quasi-particles is
considered as a perturbation of an electron-phonon system, which
was strongly coupled initially. Such consideration is applicable
if $\epsilon_F\gg\omega_0\gg t_{\bot}$, where $\epsilon_F$ is
Fermi energy, $\omega_0$ is a typical phonon energy. Both
inequalities hold for layered anisotropic HTSC, where according to
\cite{Ho} in the $c$-direction an electron is surrounded by a
large number of phonons, forming a polaron \cite{Hols} which
influences the transversal $ab$-transport weakly. The following
analytic expression was obtained in \cite{Ho} for Einsteinian
spectrum of $c$-polarized phonons in the temperature range
$T\sim\omega_0$:
\begin{equation}
\rho_c(T)\propto \rho_{ab}(T)\frac{\exp[g^2\tanh(\omega_0/4T)]}
{\sqrt{\sinh(\omega_0/2T)}},
\end{equation}
where $g$ is a parameter characterizing strength of
electron-phonon interaction, $g>1$. Fig.~4 represents our result
of comparing experimental dependences $\rho_c(T)$ (symbols) and
those calculated from (6) (solid lines). To calculate them, we
used $\rho_{ab}(T)$ data in Fig.~2; $g$ was almost the same for
all the dependences in Fig.~4: $g\approx 3$; $\omega_0$ increased
from 110~K (75~cm$^{-1}$) to 310~K (215~cm$^{-1}$) when the oxygen
content $(7-x)$ was decreased in YBa$_2$Cu$_3$O$_{7-x}$ from 6.93
to 6.53. In this connection, it seems quite natural to observe the
anomalies of optical $c$-conductivity of YBa$_2$Cu$_3$O$_{7-x}$
crystals with oxygen deficiency in this frequency
range~\cite{Tim}.

Thus, this paper represents the measurements of anisotropy of
microwave conductivity of YBa$_2$Cu$_3$O$_{7-x}$ crystal, having
varied the holes concentration $p$ in the range $0.08\leq p\leq
0.15$. Having analyzed temperature dependences of imaginary parts
of the conductivity tensor $\hat\sigma''(T)$ in the
superconducting state and the resistivity $\hat\rho(T)$ in the
normal state, we have demonstrated that the optimally doped
YBa$_2$Cu$_3$O$_{6.93}$ is a tree-dimensional anisotropic metal.
Decrease of carriers concentration in it results in the crossover
from Drude $c$-axis conductivity to the hopping one. Moreover, to
quantitatively describe the evolution of the dependences
$\sigma''_c(T)$ and $\rho_c(T)$ at varying $p$, it is necessary to
allow for the effects of strong electron-phonon interaction.

The authors acknowledge V.F.~Gantmakher and A.F.~Shevchun for
valuable discussions. This work was supported by RFBR Grants Nos.
03-02-16812, 03-02-06386, 02-02-08004.


\begin{references}

\bibitem{Tal1} J.L.~Tallon, C.~Bernhard, H.~Shaked et al., Phys. Rev. B
{\bf 51}, 12911 (1995).

\bibitem{Xie} Y.B.~Xie, Phys. Rev. B {\bf 45}, 11375 (1992).

\bibitem{Tru1} M.R.~Trunin and A.A.~Golubov, in {\it Spectroscopy of High-$T_c$
Superconductors. A Theoretical View} (Taylor and Francis, London and New
York, 2003), pp.159-233.

\bibitem{Mao}  J.~Mao, D.H.~Wu, J.L.~Peng et al.,
Phys. Rev. B {\bf 51}, 3316 (1995).

\bibitem{Srik} H.~Srikanth, Z.~Zhai, S.~Sridhar et al.,
J. Phys. Chem. Solids {\bf 59}, 2105 (1998).

\bibitem{Nef1} Yu.A.~Nefyodov, M.R.~Trunin, A.A.~Zhohov et al.,
Phys. Rev. B {\bf 67}, 144504 (2003).

\bibitem{Hos} A.~Hosseini, S.~Kamal, D.A.~Bonn et al.,
Phys. Rev. Lett. {\bf 81}, 1298 (1998).

\bibitem{Xiang} T.~Xiang, C.~Panagapoulos, and J.R.~Cooper, Int. Journ.
Mod. Phys. B {\bf 12}, 1007 (1998).

\bibitem{Nef2} Yu.A.~Nefyodov and M.R.~Trunin, Physica C {\bf 388-389}, 469 (2003).

\bibitem{Lev1} R.J.~Radtke, V.N.~Kostur, and K.~Levin, Phys. Rev. B
{\bf 53}, R522 (1995); R.J.~Radtke and K.~Levin, Physica C {\bf 250}, 282
(1995); R.J.~Rojo and K.~Levin, Phys. Rev. B {\bf 48}, 16861 (1993).

\bibitem{Legg} M.~Turlakov and A.J.~Legget, Phys. Rev. B
{\bf 63}, 064518 (2001).

\bibitem{Kum} N.~Kumar and A.M.~Jayannavar, Phys. Rev. B {\bf 45}, 5001 (1992).

\bibitem{Lar} L.B.~Ioffe, A.I.~Larkin, A.A.~Varlamov et. al., Phys. Rev. B
{\bf 47}, 8936 (1993).

\bibitem{Graf} M.J.~Graf, D.~Rainer, and J.A.~Sauls, Phys. Rev. B
{\bf 47}, 12089 (1993).

\bibitem{Ho} A.F.~Ho and A.J.~Schofield, cond-mat/0211675.

\bibitem{Lan} I.G.~Lang and Yu.A.~Firsov, Sov. Phys. JETP {\bf
16}, 1301 (1963).

\bibitem{Hols} T.~Holstein, Ann. of Phys. {\bf 8}, 343 (1959).

\bibitem{Tim} T.~Timusk and B.~Statt, Rep. Prog. Phys. {\bf 62}, 61 (1999).

\end{references}
\end{document}